\documentstyle[prl,aps,floats]{revtex}

\begin{document}
\widetext
\draft
\twocolumn[\hsize\textwidth\columnwidth\hsize\csname @twocolumnfalse\endcsname
\title{Disorder-Induced Anomalous Kinetics in the $\mbox{A}+\mbox{A}
\to \emptyset$
Reaction}

\author{
Jeong-Man Park$^{1,2}$
and 
Michael W. Deem$^1$ 
}
\address{
$^1$Chemical Engineering Department, University of
California, Los Angeles, CA  90095-1592\\
$^2$Department of Physics, The Catholic University, Seoul, Korea}

\maketitle

\begin{abstract}
We address the two-dimensional bimolecular annihilation reaction
$\mbox{A} + \mbox{A} \to \emptyset$ in the presence of random impurities.
Impurities with sufficiently long-ranged interaction energies
are known to lead to anomalous diffusion,
$\langle r^2(t) \rangle \sim t^{1-\delta}$, in the absence
of reaction.  Applying renormalization group theory
to a field theoretic description of this reaction, we find
that this disorder also leads to
anomalous kinetics in the long time limit:
$c(t) \sim t^{\delta -1}$.  This kinetics results because
the disorder forces the system into the (sub)diffusion controlled
regime, in which the kinetics must become anomalous.
\end{abstract}

]

\narrowtext

Surface reactions show a variety of complex spatial
and temporal patterns.  Simple systems, such as
oxidation of CO on single crystal Pt(110), show
surprisingly rich behavior \cite{Graham}, ranging
from spirals and standing waves to chemical
turbulence \cite{Ertl,Jakubith}.  Such behavior
results because two dimensions is the upper critical
dimension for many surface reactions, and so collective
fluctuations control the dynamics.  Toussaint and Wilczek, and
Kang and Redner first noticed the diffusion-controlled kinetics that
can arise from reactant microphase separation
\cite{Toussaint,Kang1,Kang2}.  Field theoretic techniques
were developed that rigorously showed two dimensions to be
the upper critical dimension for the annihilation
reactions $\mbox{A} + \mbox{A} \to \emptyset$ \cite{Peliti,Lee1} and
$\mbox{A} + \mbox{B} \to \emptyset$ \cite{Lee2}.  Adsorption effects
have been addressed within the field theoretic framework
 \cite{Droz}.  Real
systems, of course, possess many defects, and 
randomness in adsorption energies can lead to a variety of
phases observed at steady state \cite{Redner2}.

Defects can dramatically affect the diffusion of
particles if correlations in the
potential field are
sufficiently long ranged.  For a single ion diffusing
in disorder that is charged, the appropriate form
of the correlation function at long-wavelengths is
\cite{Deem0}
\begin{equation}
\hat \chi_{vv}(k) = \gamma/k^2 \ .
\label{3}
\end{equation}
For this model, the upper critical dimension is two.
In two dimensions and below, 
 subdiffusion
occurs \cite{Fisher,Kravtsov1}.  In this anomalous regime the
mean-square displacement increases sublinearly with time,
$\langle r^2(t) \rangle \sim t^{1-\delta}$.
The diffusion 
exponent, $1-\delta$,  depends on the strength of
disorder; it
can be found exactly
\cite{Kravtsov2,Bouchaud1,Bouchaud2,Honkonen1,Honkonen2,Derkachov1,Derkachov2}.

These same defects can have an effect on reacting species.
We consider species that diffuse in this quenched,
random potential and which also react according to the law
$\mbox{A}+\mbox{A} \to \emptyset$.
The diffusing species, however, do not interact in any
other way.

The type of physical system that we have in mind
is a reaction that occurs on the surface of a crystalline
ionic lattice.  The substrate lattice has dislocation line
pairs, which form line vacancies or line interstitials.
These defects are immobile and generate a random, quenched
electrostatic potential on the surface, represented by
Eq.\ (\ref{3}).  On the surface, the reaction $\mbox{A}+\mbox{A} \to \emptyset$
occurs, where $\mbox{A}$ is a reactive ion.
The $1/r$ interaction between the ions is technically
irrelevant, and so it can be ignored.
So as to maintain surface charge neutrality, however, we may wish to include
additional counter ions, which can desorb.

Alternatively, our results can be viewed as approximately describing
the annealing of two-dimensional topological defects, such as
line dislocations in solids. 
Topological defects have
logarithmic interactions, due to a slowly decaying elastic strain field
(see \cite{nelsonII} for a review).
If some of the defects were pinned and unreactive, the remaining
defects would diffuse and combine in the quenched, random potential
described by Eq.\ (\ref{3}).
The logarithmic interaction between the topological defects
prevents microphase separation, and so the kinetics of this two-species
reaction is similar to that of 
the $\mbox{A}+\mbox{A}$ reaction \cite{Toussaint}.

In this Rapid Communication
we analyze a field theoretic formulation of the
two-dimensional annihilation
reaction $\mbox{A} + \mbox{A} \to \emptyset$ in the
presence of quenched disorder.  The concentration of
$\mbox{A}$ is initially Poissonian, with average density $n_0$.
No creation or unimolecular annihilation is allowed.  We
search for the asymptotic decay law at long times using
renormalization group (RG) theory.  A field theory is derived
by identifying a master equation, writing the master equation
in terms of creation and annihilation operators, and
using the coherent state representation \cite{Peliti,Lee1}.  The
random potential is incorporated with the replica trick \cite{Kravtsov1}.
The concentration of $\mbox{A}$ at time $t$, $c({\bf x},t)$, is given by
\begin{equation}
c({\bf x},t) = \lim_{N \to 0} \langle a({\bf x},t) \rangle \ ,
\label{1}
\end{equation}
where the average is taken with respect to  $\exp(-S)$, with
the action $S = S_0 + S_1 + S_2 + S_3$,
\begin{eqnarray}
S_0 &=& \int d^d {\bf x} \int_0^{t_f} d t
 \bar a_\alpha({\bf x},t) \left[
\partial_t - D \nabla^2 + \delta(t)
 \right]
 a_\alpha({\bf x},t)
 \nonumber \\ 
S_1 &=& 
\int d^d {\bf x} \int_0^{t_f} d t \bigg[
\lambda_1 \bar a_\alpha({\bf x},t)
 a_\alpha^2({\bf x},t)
\nonumber \\ ~~~~~
&&+\lambda_2 \bar a_\alpha^2({\bf x},t) 
 a_\alpha^2({\bf x},t) 
\bigg]
\nonumber \\ 
S_2 &=& -n_0 \int d^d {\bf x} \bar a_\alpha({\bf x},t)
\nonumber \\ 
S_3 &=& \frac{\beta^2 D^2}{2}
\int d t_1 d t_2 \int_{{\bf k}_1 {\bf k}_2 {\bf k}_3 {\bf k}_4}
\nonumber \\ 
&& \times (2 \pi)^d \delta({\bf k}_1+{\bf k}_2+{\bf k}_3+{\bf k}_4)
\nonumber \\ &&\times 
\hat{\bar a}_{\alpha_1}({\bf k}_1, t_1)
\hat{     a}_{\alpha_1}({\bf k}_2, t_1)
\hat{\bar a}_{\alpha_2}({\bf k}_3, t_2)
\hat{     a}_{\alpha_2}({\bf k}_4, t_2)
\nonumber \\ &&\times 
{\bf k}_1 \cdot ({\bf k}_1+{\bf k}_2)
{\bf k}_3 \cdot ({\bf k}_1+{\bf k}_2)
\hat\chi_{vv}(\vert {\bf k}_1+{\bf k}_2\vert)
\ .
\label{2}
\end{eqnarray}
Summation is implied over replica indices.  The notation
$\int_{\bf k}$ stands for $\int d^d {\bf k} / (2 \pi)^d$.
The upper time limit in the action is arbitrary as long as 
$t_f \ge t$.  The term $S_0$ represents simple diffusion,
without an external potential.  The delta function, often
left out by convention, enforces the initial condition on the free
field propagator.  The term $S_1$ comes from the reaction terms.
The parameters are related to the conventional reaction rate,
$k$, by $\lambda_1 = 2 \lambda_2  = k$.  The term $S_2$ comes
from the random, Poissonian initial condition.  The term $S_3$
comes from averaging the concentration over the random potential.
The potential is assumed to be Gaussian, with zero mean and
correlation function given by Eq.\ (\ref{3}).
The inverse temperature is given by $\beta = 1/(k_{\rm B} T)$.

An approximate solution to Eq.\ (\ref{1}) can be derived by
a saddle point approximation to the action (\ref{2}).  The
result, before the average over the random potential is taken,
is 
\begin{eqnarray}
\partial_t c_v &=& D \nabla^2 c_v + \beta D \nabla \cdot
\left( c_v \nabla v\right) - \lambda_1 c_v^2
\nonumber \\ 
c_v({\bf x}, 0) &=& n_0 \ .
\label{4}
\end{eqnarray}
For constant potential, this equation has the solution 
$c_v({\bf x}, t) = 1/(\lambda_1 t + 1/n_0)$.

An alternative, exact expression for the concentration can
be derived by performing a Hubbard-Stratonovich transformation
on (\ref{2}) and integrating out the field $\bar a$:
\begin{eqnarray}
\partial_t c_{\eta v} &=& D \nabla^2 c_{\eta v} + \beta D \nabla \cdot
\left( c_{\eta v} \nabla v\right) - \lambda_1 c_{\eta v}^2
\nonumber \\ 
&& +i \eta c_{\eta v}
\nonumber \\ 
c_{\eta v}({\bf x}, 0) &=& n_0 \ ,
\label{5}
\end{eqnarray}
where the real, Gaussian, random field $\eta$ has zero mean and
variance $\langle \eta({\bf x},t) \eta({\bf x}',t') \rangle 
= 2 \lambda_2 \delta({\bf x}- {\bf x}') \delta(t-t')$.
The physical concentration is given by averaging the
solution over the random field $\eta$.  From this
representation, we see that mean field theory is exact when
$\lambda_2$ = 0.  It is for this reason that diagrams to
all orders in $\lambda_1$ can be summed in this limit
(see, for example, \cite{Lee1}).

We apply renormalization group theory to the action (\ref{2}) to
take into account the effects of nonzero $\lambda_2$.
Time ordering prevents a term of the form $\bar a a$ from
being generated, and all other relevant terms are already present
in Eq.\ (\ref{2}).  The limit $N \to 0$ turns out not to matter,
as time ordering cancels the same terms that this limit does.
We follow the flow equations until $\lambda_2$ is
small enough so that we have entered the perturbative
regime.  In this regime, we can match the flow equations with
the solution to Eq.\ (\ref{4}).  We integrate over momenta
in the range $\Lambda/b < k < \Lambda$ and then rescale the
fields by 
$\hat{\bar a}'(b {\bf k},b^{-z} t) = 
\hat{\bar a}( {\bf k},t) / \bar \alpha$
 and
$\hat { a}'(b {\bf k},b^{-z} t) = 
\hat { a}( {\bf k},t) /  \alpha$.
To achieve a fixed point, and to keep the time derivative in $S_0$ 
constant, we set $\alpha =1, \bar \alpha = b^d$.  We determine the
dynamical exponent by requiring that the diffusion 
coefficient remain unchanged.  The flow equations in two dimensions,
to one loop order, then become
\begin{eqnarray}
\frac{d \ln n_0}{d l} &=& 2
\nonumber \\
\frac{d \ln \lambda_1}{d l} &=& - 
   \frac{\lambda_2}{2 \pi D} + \frac{3 \beta^2 \gamma}{4 \pi}
\nonumber \\
\frac{d \ln \lambda_2}{d l} &=& - 
   \frac{\lambda_2}{2 \pi D} + \frac{3 \beta^2 \gamma}{4 \pi}
\nonumber \\
\frac{d \beta^2 \gamma}{d l} &=& 0
\ .
\label{6}
\end{eqnarray}
The dynamical exponent is given by
\begin{equation}
z = 2 + \frac{\beta^2 \gamma}{4 \pi} \ .
\end{equation}

These flow equations reproduce known anomalous scaling in the
cases of no reaction or no disorder.  They are integrated to a
time such that
\begin{equation}
t(l^*) = t e^{-\int_0^{l^*} z(l) dl } = t_0 \ .
\end{equation}
The matching time, $t_0$, is chosen to be on the order of
 $4 \pi^2/(\Lambda^2 D)$
 so as to
be within the range of validity of both RG scaling and mean field
theory.  Mean field theory is a good approximation at this time
because $\lambda_2(l^*)$ is small.  (An expansion in $\lambda_2$
generates an expansion in $t_0$ that leads to subdominant
scaling.)  For short times, particles see only the local value of
the potential, and the effective diffusivity is given by the
bare value.  (An expansion in $\beta$ generates an expansion
in $t_0$ that leads to subdominant scaling.)  We can,
therefore, ignore the potential term in Eq.\ (\ref{4}) for
short times.  Matching with mean field theory, we find
the mean square displacement is given by
\begin{equation}
\left\langle r^2(t(l^*), l^*) \right\rangle = 4 D t(l^*)
\label{9}
\end{equation}
in the absence of reaction,
and the concentration profile is given by
\begin{equation}
c(t(l^*), l^*) = \frac{1}{1/n_0(l^*) + \lambda_1(l^*) t(l^*)}
\label{10}
\end{equation}
in the absence of random forces.
The observed values are related by scaling:
\begin{eqnarray}
\left\langle r^2(t) \right\rangle &=&
e^{2 l} \left\langle r^2(t(l), l) \right\rangle 
\nonumber \\
c(t) &=& e^{-2 l} c(t(l), l) \ .
\label{11}
\end{eqnarray}

With no reaction, the one loop flow equations prove to be exact
\cite{Kravtsov2,Bouchaud1,Bouchaud2,Honkonen1,Honkonen2,Derkachov1,Derkachov2}.
The disorder strength is not renormalized to any order.  The
dynamical exponent is greater than two, and so we have
subdiffusive behavior.  Matching at $l = l^*$, we find the
mean square displacement at long times is given by
\begin{equation}
\left\langle r^2(t) \right\rangle
    \sim 4 D t \left( \frac{t}{t_0} \right)^{-\delta} \ ,
\label{12}
\end{equation}
with
\begin{equation}
\delta = \left[ 1 + \frac{8 \pi}{\beta^2 \gamma} \right]^{-1} \ .
\label{13}
\end{equation}
We have a continuously variable exponent that depends on the strength
of disorder.  The prefactor is nonuniversal, since the matching
time $t_0$ explicitly enters.

With no external potential, the flow equations become asymptotically
exact at long times \cite{Peliti,Lee1}.
  The rate constants decay slowly to zero, with the
ratio $\lambda_2(l)/\lambda_1(l)$ remaining constant.  The
dynamical exponent is two, since there is no field renormalization.
Matching with mean field theory at $l=l^*$, we find for long
times
\begin{equation}
c(t) \sim \frac{\ln(t/t_0)}{8 \pi D t} \ .
\label{14}
\end{equation}
We see the characteristic universal logarithmic
correction to the mean-field kinetic equations.  There is
a subdominant, nonuniversal $1/t$ decay.

The presense of disorder dramatically affects the reaction.  The
rate constants no longer decay to zero, but to a nonzero
fixed point.  The ratio $\lambda_2(l)/\lambda_1(l)$
again remains constant.  To one loop, we find the reaction
terms do not renormalize the diffusivity, disorder strength, or
dynamical exponent.  Matching with mean field theory at $l = l^*$, we
find
\begin{equation}
c(t) \sim \frac{1}{\lambda_1^* t} \left( \frac{t}{t_0} \right)^\delta \ ,
\label{15}
\end{equation}
with the fixed point given from Eq.\ (\ref{6}) as
\begin{equation}
\lambda_1^* = 3 D \beta^2 \gamma \ .
\label{16}
\end{equation}
We have, therefore, a nonuniversal decay exponent that
depends, again, on the strength of disorder.  The exponent
is precisely the inverse of that for the pure diffusion case, Eq.\ (\ref{12}).
The prefactor
is also nonuniversal, since the matching time $t_0$ again
enters.

We see that to one loop, the reaction terms leave the diffusion
process unaffected.  Physically, the disorder traps diffusing
species in wells of deep potential energy.  Normal diffusion
between these wells leads to the observed, overall diffusivity.
Progressively deeper wells are encountered at longer times, and so
anomalous diffusion occurs.  
  For diffusion limited reactions, the effective reaction
rate is proportional to the effective diffusion constant.  Since we
have $D_{\rm eff}(t) \sim D (t/t_0)^{-\delta}$, we should expect
$k_{\rm eff} \propto (t/t_0)^{-\delta}$.  This is exactly the
behavior observed in Eq.\ (\ref{15}).

When two reactants are attracted to the same
well, reaction almost always occurs, since the escape time is much
greater than the reaction time.  In other words, the reaction becomes
(sub)diffusion limited, and the bare reaction rate should not enter the 
decay law.  Three species entering a well is much less probable at low
densities than two, and so this case should be irrelevant at
long times.
Reactants can encounter each other outside the wells. Reaction
outside the wells would show up as renormalization of the disorder
strength and dynamical exponent in a two-loop calculation.
We have checked that these terms are not generated to two-loop order.
So  the dynamical exponent
is unchanged by the reaction, and the bare reaction rate is not
present in the decay law.  From this argument, we expect the 
dynamical exponent and the flow equation for the disorder to
remain unchanged in higher order calculations.

The diffusion terms, on the other hand,  do affect the reaction process.
Since the reaction becomes diffusion limited, the diffusion terms
actually control the reaction process.  For this reason, we find a
nonzero fixed point for the effective reaction rate.  The
effective reaction rate should be a function of the strength of
disorder, since it is related to the rate at which species
diffuse between wells.  The rate (at $l = l^*$) is proportional
to the effective diffusion coefficient divided by a characteristic
distance squared.  This is exactly the behavior that we observe in Eq.\ 
(\ref{16}), with $\beta^2 \gamma$ an inverse distance squared.
Higher order calculations may, of course, lead to modification of the
numerical value of the fixed point for large $\beta^2 \gamma$.

This research was supported by the National Science Foundation
though grants CHE--9705165 and CTS--9702403.

\bibliography{react}

\end{document}